\documentclass[onecolumn,nofootinbib]{revtex4}
\usepackage{graphicx}
\usepackage{latexsym}
\newcommand{\be}{\begin{equation}}
\newcommand{\ee}{\end{equation}}
\newcommand{\bea}{\begin{eqnarray}}
\newcommand{\eea}{\end{eqnarray}}

\begin{document}


\title{Probing for dynamics of dark energy and curvature of universe with latest cosmological observations}

\author{Gong-Bo Zhao${}^a$, Jun-Qing Xia${}^a$, Hong Li${}^b$, Charling Tao${}^c$, Jean-Marc Virey${}^d$,
 Zong-Hong Zhu${}^e$, and Xinmin Zhang${}^a$ }

\affiliation{${}^a$Institute of High Energy Physics, Chinese
Academy of Science, P.O. Box 918-4, Beijing 100049, P. R. China}

\affiliation{${}^b$Department of Astronomy, School of Physics,
Peking University, Beijing, 100871, P. R. China}

\affiliation{${}^c$Centre de Physique des Particules de Marseille, CNRS/IN2P3-Luminy and Universit\'e de la M\'editerran\'ee,
Case 907, F-13288 Marseille Cedex 9, France}

\affiliation{${}^d$Centre de Physique Th\'eorique\thanks{``Centre de Physique Th\'eorique'' is UMR 6207 - ``Unit\'e Mixte
de Recherche'' of CNRS and of the Universities ``de Provence'',
``de la M\'editerran\'ee'' and ``du Sud Toulon-Var''- Laboratory
affiliated to FRUMAM (FR 2291).}, CNRS-Luminy and Universit\'e de Provence, Case 907,
F-13288 Marseille Cedex 9, France.}

\affiliation{${}^e$Department of Astronomy, Beijing Normal University, Beijing 100875, P. R. China}

\begin{abstract}

We use the newly released 182 Type Ia supernova data combined with
the third-year Wilkinson Microwave Anisotropic Probe data (WMAP3)
and large scale structure (LSS) information including SDSS and
2dFGRS to constrain the dark energy equation of state (EoS) as
well as the curvature of universe $\Omega_K$. Using the full
dataset of Cosmic Microwave Background (CMB) and LSS rather than
the shift parameter and linear growth factor, we make a Markov
Chain Monte Carlo (MCMC) global fit, while paying particular
attention to the dark energy perturbation. Parameterizing the EoS
as $w_{DE}(a) = w_{0} + w_{1}(1-a)$, we find the best fit of
($w_0,w_1$) is ($-1.053,0.944$) and for $w_{DE}(a) = w_{0} +
w_{1}\sin(\frac{3}{2}\pi \ln(a))$, the best fit for ($w_0,w_1$) is
($-1.614,-1.046$). We find that a flat universe is a good
approximation, namely, $|\Omega_K|>0.06$ has been excluded by
2$\sigma$ yet the inclusion of $\Omega_K$ can affect the
measurement of DE parameters owing to their correlation and the
present systematic effects of SNIa measurements.

\end{abstract}

\maketitle

\section{Introduction}

Dark energy (DE), the very power to drive universe's acceleration,
is one of the most important issues in modern cosmology. Its
existence was firstly revealed by the measurement of the
relationship between redshift $z$ and luminosity distance $d_L$ of
Type Ia supernova (SN Ia)\cite{SN98}. Dark Energy encodes its
mystery in its equation of state (EoS) defined as the ratio of
pressure over energy density thus DE models can be classified in
terms of EoS\cite{DEclassify}.

The simplest candidate of dark energy is the cosmological constant
(CC) whose EoS remains $-1$. Favored by current astronomical
observations as it is, CC suffers from severe theoretical
drawbacks such as the fine-tunning and coincidence
problem\cite{CCproblem}. Alternative DE models with rolling scalar
field, such as quintessence\cite{quintessence},
phantom\cite{phantom}, k-essence\cite{kessence}, etc have been studied. The EoS of
these models varies with cosmic time either above $-1$ or below
$-1$ during evolution but the statement of ``No-Go" theorem
forbids it to cross the $-1$ boundary\cite{perturbation1}. Models
where gravity is modified can also give these observed effects.

Given our ignorance of the nature of dark energy, constraining the
evolution of DE the EoS by cosmological observations is of great
significance. Various methods have been used to constrain DE
including parametric
fitting\cite{perturbation2,ourmcfit,othersfit}, non-parametric
reconstruction\cite{reconstruct}, \emph{etc}. Interestingly, there
exists some hint that the EoS of DE has crossed over $-1$ at least
once from current
observations\cite{ourmcfit,reconstruct,Barger,hongli},
which greatly challenges the above mentioned dark energy models,
albeit the evidence is still marred by systematic effects.
Quintom, whose EoS can smoothly cross $-1$\cite{quintom}, has
attracted a lot of attention in the literature since its
invention\cite{study4quintom}. There have been many efforts in
quintom model building, for example, double-scalar-field
realization\cite{quintom,2fields}, a single scalar field with high
derivative\cite{HD}, vector fields\cite{vector} and so forth.

A special and interesting example of quintom is Oscillating
Quintom, whose EoS oscillates with time and crosses $-1$ many
times. The oscillating behavior in the EoS leads to oscillations
in the Hubble constant and a recurrent universe. Oscillating
Quintom is physically well motivated, since this scenario, to some
extent, unifies early inflation and the current acceleration of
the universe\cite{oscq_1}. In Ref.\cite{oscq_2}, we have presented
some preliminary studies on oscillating quintom.

The nature of dark energy is a dominant factor of the fate of our
universe.
Another critical point is the curvature, $\Omega_K$.  neutrino masses, which
probably exists, albeit small, and the curvature, $\Omega_K$.
We concentrate, in
this paper, on the correlation between $\Omega_K$ and the dark
energy parameters.\footnote{Other cosmological parameters also affect
the probing of dark energy, such as neutrino mass\cite{neutrino} and inflationary parameters\cite{inflation}.}

We use the newly released 182 SN Ia ``Gold sample"
(SN182)\cite{sn182} combined with
WMAP3\cite{wmap3}\footnote{Available at
http://lambda.gsfc.nasa.gov/product/map/current/} and LSS
information to constrain the evolution of the DE EoS and the
curvature of universe. For DE EoS, we choose two parameterizations
as in (\ref{EOS1}) and (\ref{EOS2}) and we will address our
motivation for such a choice in the next section. In our study, we
treat the curvature of the universe $\Omega_K$ as a free parameter
rather than simply assuming a flat universe and make a Markov
Chain Monte Carlo (MCMC) global fit based on Bayesian statistics.
Paying particular attention to the dark energy perturbation
especially when EoS crosses $-1$\cite{perturbation2}, we find the
latest observations mildly favor quintom model however
$\Lambda$CDM remains a good fit. We have also found that the
inclusion of $\Omega_K$ can affect the determination of DE
parameters significantly due to their correlation.




We structure this paper as follows: after this introductory part,
we propose our method and define the data set used in section II.
The in section III we present our results and end up with
discussion and comments.

\section{Method and data}

To study the dynamical behavior of dark energy, we choose two kinds of
parametrization of dark energy equation of state:\\
\noindent I)
\begin{equation}
\label{EOS1} w_{DE}(a) = w_{0} + w_{1}(1-a)
\end{equation}

\noindent II)
\begin{equation}
\label{EOS2} w_{DE}(a) = w_{0} + w_{1}\sin(w_2\ln(a))
\end{equation}
where $a$ is the scale factor, $w_0$ denotes the EoS at present
epoch and $w_{1}$ and $w_2$ characterize the time evolution of DE.
Parametrization I) is the most popular in literature since $w_{1}$
simply equals to $-dw_{DE}(a)/da$, which is the time derivative of
$w_{DE}(a)$\cite{Linderpara}. Thus it is straightforward to study
the dynamical behavior of DE. The physical motivation of
parametrization II) is oscillating quintom. From (\ref{EOS2}) we
can see at low redshift, II) takes a form similar to I). At medium
and high redshift, the EoS keeps oscillating.

From the latest SN Ia paper\cite{sn182}, one can find some hint of
oscillating behavior of the EoS in their FIG.10 where they use a
quartic polynomial fit. Our sine function has the advantage of
preserving the oscillating feature of the EoS at high redshift
measured by the CMB data. For simplicity and focus on the study at
lower redshift, we set $w_2$ to be $\frac{3}{2}\pi$ in order to
allow the EoS to evolve more than one period within the redshift
range of 0 to 2 where SN data are most robust.

When using the MCMC global fitting strategy to constrain
cosmological parameters, it is crucial to include dark energy
perturbation. This issue has been realized by many researchers
including the WMAP
group\cite{perturbation1,perturbation2,wmap3,alewis}. However one
cannot handle the dark energy perturbation when the parameterized
EoS crosses $-1$ based on quintessence, phantom,k-essence and
other non-crossing models. By virtue of quintom, the perturbation
at the crossing points is continuous, thus we have proposed a
sophisticated technique to treat dark energy perturbation in the
whole parameter space, say, EoS$>-1$, $<-1$ and at the crossing
pivots. For details of this method, we refer the readers to our
previous companion paper \cite{perturbation1,perturbation2}.

In this study, we have modified the publicly available Markov
Chain Monte Carlo package CAMB/CosmoMC\cite{CosmoMC} to include
the dark energy perturbation when the equation of state crosses
$-1$.

The dark energy EoS and curvature of the universe $\Omega_K$ can
affect the determination of the geometry of our universe thus DE
parameters are correlated with $\Omega_K$. Therefore, in our
study, we relax the curvature of universe $\Omega_K$ as a free
parameter rather than simply assuming a flat universe. We assume
purely adiabatic initial conditions and set our most general
parameter space as:
\begin{equation}
\label{parameter} {\bf P} \equiv (\omega_{b}, \omega_{c},
\Theta_{s}, \tau, \Omega_K, w_{0}, w_{1}, n_{s},
\ln(10^{10}A_{s}))
\end{equation}
where $\omega_{b}\equiv\Omega_{b}h^{2}$ and
$\omega_{c}\equiv\Omega_{c}h^{2}$ are the physical baryon and Cold
Dark Matter densities relative to the critical density,
$\Theta_{s}$ is the ratio (multiplied by 100) of the sound horizon
to the angular diameter distance at decoupling, $\tau$ is the
optical depth to re-ionization,
$\Omega_K\equiv1-\Omega_m-\Omega_{DE}$ is the spatial curvature,
$w_{0}, w_{1}$ portray the dynamical feature of dark energy as
illustrated in (\ref{EOS1}) and (\ref{EOS2}). $A_{s}$ and $n_{s}$
characterize the primordial scalar power spectrum. For the pivot
of the primordial spectrum we set $k_{s0}=0.05$Mpc$^{-1}$.
Furthermore, we make use of the Hubble Space Telescope (HST)
measurement of the Hubble parameter $H_{0}\equiv
100$h~km~s$^{-1}$~Mpc$^{-1}$\cite{Hubble} by multiplying the
likelihood by a Gaussian likelihood function centered around
$h=0.72$ and with a standard deviation $\sigma=0.08$. We also
impose a weak Gaussian prior on the baryon density
$\Omega_{b}h^{2}=0.022\pm0.002$ (1 $\sigma$) from Big Bang
Nucleosynthesis\cite{BBN}. The weak priors we take are as follows:
$\tau<0.8$, $0.5<n_{s}<1.5$, $-0.3<\Omega_K<0.3$, $-3<w_0<3$,
$-5<w_1<5$ \footnote{We set the prior of $w_0$ and $w_1$ broad
enough to ensure the EoS can evolve in the whole parameter space.}
and a cosmic age tophat prior 10 Gyr $< t_0 <$ 20 Gyr.

In our calculations, we have taken the total likelihood to be the
products of the separate likelihoods ${\bf \cal{L}}$ of CMB, LSS
and SNIa. In other words, defining $\chi^2 \equiv -2 \log {\bf
\cal{L}}$, we get \be\label{chi2} \chi^2_{total} = \chi^2_{CMB} +
\chi^2_{LSS} + \chi^2_{SNIa} ~ . \ee If the likelihood function is
exactly Gaussian, $\chi^2$ coincides with the usual definition of
$\chi^2$ up to an additive constant corresponding to the logarithm
of the normalization factor of ${\cal L}$. In the calculation of
the likelihood from SNIa we have marginalized over the nuisance
parameter  \cite{DiPietro:2002cz}. The supernova data we use are
the ``gold'' set of 182 SNIa recently published by Riess $et$ $al$
in Ref.\cite{sn182}. In the computation of CMB we have used the
full dataset of the WMAP3 data with the routine for computing the
likelihood supplied by the WMAP team
 \cite{wmap3}.
 For LSS information, we have used the 3D power
spectrum of galaxies from the SDSS \cite{Tegmark:2003uf} and
2dFGRS\cite{Cole:2005sx}. To be conservative but more robust, in
the fittings to the 3D power spectrum of galaxies from the SDSS,
 we have used the first 14 bins only, which
are supposed to be well within the linear regime  \cite{sdssfit}.

For each regular calculation, we run six independent chains
comprising $150,000-300,000$ chain elements and spend thousands of
CPU hours to calculate on a cluster. The average acceptance rate
is about $40\%$. We discard the first $30\%$ chain elements to be
the "burn-in" process, test the convergence of the chains by
Gelman and Rubin criteria\cite{R-1} and find $R-1$ of order
$0.01$, which is more conservative than the recommended value
$R-1<0.1$.

\section{Results}

\begin{table}
TABLE 1. Constraints of dark energy equation of state and some
background parameters when relaxing $\Omega_K$ as a free parameter
(Left panel) and assuming a flat universe (Right). For each case,
we consider two forms of parametrization of dark energy EoS:
Linear ($w(a)=w_0+w_1(1-a)$) and Oscillating
($w(a)=w_0+w_1\sin(w_2\ln(a))$, set $w_2=\frac{3}{2}\pi$, see text
for explanation). Best fit models, which give the minimum
$\chi^2$, and the marginalized $2\sigma$ errors are shown. All
these constraints are from data combination of WMAP3 + SN182 +
SDSS + 2dFGRS.
\begin{center}
\begin{tabular}{c||c|c|c|c}

\hline
\hline

&\multicolumn{4}{c}{$w(a)=w_0+w_1(1-a)$}  \\
\hline

&\multicolumn{2}{c|}{$\Omega_K$ free} &\multicolumn{2}{c}{$\Omega_K$=0} \\


&\multicolumn{1}{c}{Best fit} &\multicolumn{1}{c|}{$2\sigma$}
&\multicolumn{1}{c}{Best fit} &\multicolumn{1}{c}{$2\sigma$}
\\


$\Omega_K$ & $-0.015$ & $[-0.058,0.028]$ & set to 0 & set to 0\\

$w_0$ & $-1.053$ & $[-1.441,-0.615]$ & $-1.149$& $[-1.269,-0.606]$\\
$w_1$ & $0.944$ & $[-1.983,1.223]$ & $1.017$ & $[-1.078,1.163]$\\
$\Omega_m$ & $0.282$ & $[0.238,0.423]$ & 0.291 & $[0.246,0.332]$\\

\hline
\hline
\end{tabular}
\end{center}
\vspace{0.5cm}
\begin{center}
\begin{tabular}{c||c|c|c|c}

\hline
\hline

&\multicolumn{4}{c}{$w(a)=w_0+w_1\sin(w_2\ln(a))$}  \\
\hline

&\multicolumn{2}{c|}{$\Omega_K$ free} &\multicolumn{2}{c}{$\Omega_K$=0} \\


&\multicolumn{1}{c}{Best fit} &\multicolumn{1}{c|}{$2\sigma$}
&\multicolumn{1}{c}{Best fit} &\multicolumn{1}{c}{$2\sigma$}
\\


$\Omega_K$ & $-0.012$ & $[-0.051,0.015]$ & set to 0 & set to 0\\

$w_0$ & $-1.614$ & $[-2.720,-0.660]$ & $-1.149$ & $[-2.454,-0.593]$\\
$w_1$ & $-1.046$ & $[-2.591,0.557]$ & $-0.525$ & $[-2.508,0.518]$\\
$\Omega_m$ & $0.280$ & $[0.240,0.393]$ & $0.276$& $[0.230,0.318]$\\

\hline
\hline
\end{tabular}
\end{center}

\end{table}

We summarize our main results in Table I. For all the combined
data (SN182+WMAP3+SDSS+2dFGRS), we find that the flat universe is
a good fit since the best fit value of $\Omega_K$ is $-0.015$ and
$-0.012$ for DE parametrization I) and II) respectively and
$|\Omega_K|>0.06$ has been excluded for more than 2$\sigma$ for
both DE parameterizations. This can be seen graphically in FIG.1.
From these 2-D contour plots of energy density of dark energy and
matter from different data combination and different DE
parameterizations, we find that the data of supernova-only favor a
non-flat universe, however when CMB and LSS data are combined, a
flat universe is preferred.

For parametrization I), we find the best fit value of ($w_0,w_1$) to be ($-1.149,1.107$) for a flat
universe. When $\Omega_K$ is freely relaxed, the best fit value of ($w_0,w_1$) is changed to
 ($-1.053,0.944$) and the error bars of nearly all the cosmological parameters have been enlarged.
We find dark energy models whose EoS can cross $-1$ are mildly favored.

The 1-D posterior distribution of $w_0,w_1,\Omega_K$ and their 2-D correlation are shown in FIG.2.
In the $w_0-\Omega_K$ and $w_1-\Omega_K$ panel, we find interesting correlation among curvature
and DE parameters. This is expected since $\Omega_K$ can contribute to luminosity distance $d_L$ via:
\begin{eqnarray}
\label{lumdis}
d_{\rm L}(z)=\frac{1+z}{H_0\sqrt{|\Omega_{k}|}} {\rm
sinn}\left[\sqrt{|\Omega_{k}|}\int_0^z
\frac{dz'}{E(z')}\right],
\end{eqnarray}

\bea
 E(z)\equiv\frac{H(z)}{H_0}=[\Omega_m(1+z)^3+\Omega_{DE}~\exp\left(3\int_0^{z}\frac{1+w(z')}{1+z'}dz'\right) +\Omega_K(1+z)^2]^{1/2}~. \eea

where ${\rm sinn}(\sqrt{|k|}x)/\sqrt{|k|}=\sin(x)$, $x$,
$\sinh(x)$ if $k=1$, 0, $-1$. Furthermore, $\Omega_K$ can modify
the angular diameter distance to last scattering surface and the
transfer function, which leaves imprints on the CMB and matter
power spectrum.

In the $w_0-w_1$ panel of FIG.2, the parameter space has been
divided into four parts. The upper right and lower left parts
denote for $w>-1$ and $w<-1$, the regions for quintessence and
phantom models respectively. The other two parts dubbed ``Quintom
A(B)" represent models whose EoS can cross -1 during evolution.
Quintom A crosses $-1$ from upside down while Quintom B transits
$-1$ from the opposite direction. We also plot the results when
assuming a flat universe for comparison. We see the best fit model
is in the region of Quintom A and the $\Lambda$CDM(the intersect
of two dot dashed lines) is still a good fit. Moreover, relaxing
$\Omega_K$ enlarges the $w_0-w_1$ contour as expected.

For parametrization II), again we find a small absolute value of
$\Omega_K$. The best fit models can cross -1 twice in the
evolution. In FIG.3, we show the correlation between $\Omega_K$
and dark energy parameters, and we find the Quintom model whose
EoS crosses -1 during evolution is preferred.

We can see the dynamics of dark energy more clearly from FIG.4. We show the best fit model and $2\sigma$
errors of $w(z)$ for the case of flat universe and relaxing $\Omega_K$ as a free parameter for the two DE
parameterizations. The best fit model of each case is quintom-like.

\section{Summary and Discussions}

In this paper we investigate the dynamics of dark energy and
curvature of universe from the data of newly released 182
supernova data combined with CMB and LSS information. Rather than
assuming a flat universe, we relax $\Omega_K$ as a free parameter
and make a MCMC global fit to measure the dark energy parameters
as well as the curvature of universe. We find the model whose EoS
can cross over $-1$ is favored for two dark energy
parameterizations we considered in this work, albeit the
$\Lambda$CDM model remains a good fit. A flat universe is
preferred, namely, the best fit value of $|\Omega_K|$ is smaller
than 0.015 for the two DE parameterizations and $|\Omega_K|>0.06$
has been excluded by more than $2\sigma$. However, the correlation
among dark energy parameters and $\Omega_K$ might not be
neglected. Freeing $\Omega_K$ enlarges the contours and even
modifies the best fit value of dark energy parameters. For
example, for parametrization II), relaxing $\Omega_K$ has changed
the best fit value of ($w_0,w_1$) from ($-1.149,-0.525$) to
($-1.614,-1.046$). This is because adding $\Omega_K$ can reduce
the total $\chi^2$ by 1.2 thus the global minimum moves to a
deeper point.

For CMB information, we use the full dataset of WMAP3, rather than
the shift parameter. The shift parameter has the advantage of easy
implementaton and much shorter calculation time, but it is not
really ``model independent". It is derived from a fiducial
$\Lambda$CDM model, thus it might lead to biased results, if one
fits to models departing significantly from the fiducial model.
Another drawback is that the shift parameter can merely offer part
of the CMB information related to background parameters, thus it
cannot constrain the perturbated dark energy.

We will have a deeper and deeper understanding of dark energy with
the accumulation of high quality cosmological observation
especially for supernova data, such as future SNAP, ESSENCE, etc.
To be bias-free as much as possible, it is better to add
$\Omega_K$ into the global analysis, use full datasets of CMB and
carefully treat dark energy perturbation.

{\it Acknowledgement. --- }
 We acknowledge the use of the Legacy Archive
for Microwave Background Data Analysis (LAMBDA). Support for
LAMBDA is provided by the NASA Office of Space Science. We have
performed our numerical analysis on the Shanghai Supercomputer
Center(SSC). We thank Bo Feng, Mingzhe Li, Weidong Li, Pei-Hong
Gu, Xiao-Jun Bi, David Polarski and Andr$\acute{e}$ Tilquin for
helpful discussions. This work is supported in part by National
Natural Science Foundation of China under Grant Nos. 90303004,
10533010 and 19925523.

\pagebreak

\begin{figure}[htbp]
\begin{center}
\includegraphics[scale=1]{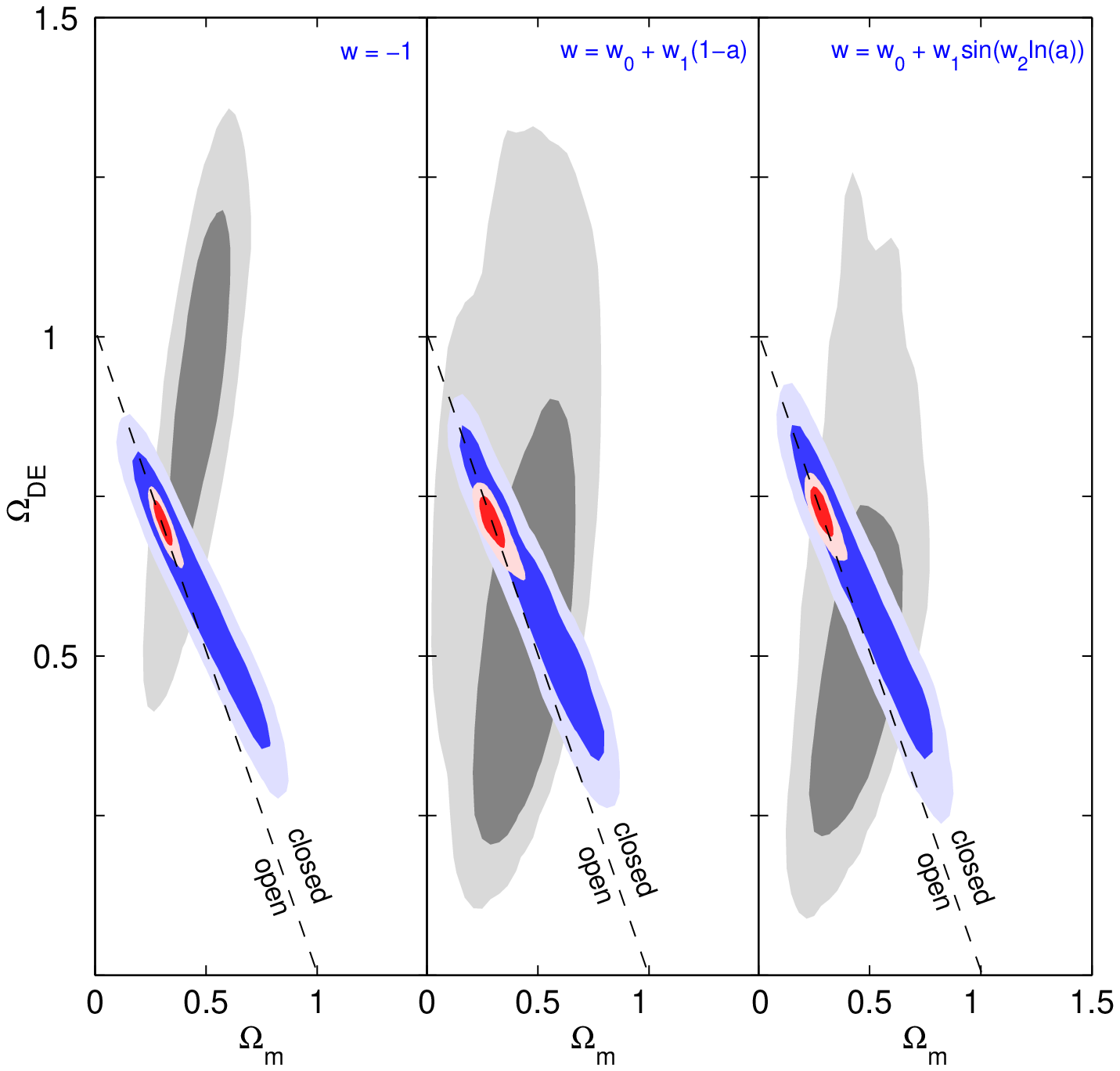}

\caption{ 2-D contour plots of energy density of matter and dark
energy using latest astronomical observations. Grey: SN182; Blue:
WMAP3; Red: WMAP3 + SN182 + SDSS + 2dFGRS. The dark and light
shaded regions stand for 68$\%$ and 95$\%$ C.L. respectively.
Different dynamical behaviors of dark energy have been considered.
Left: vacuum energy ($w = -1$ forever); Middle: linearly growing
($w(a)=w_0+w_1(1-a)$); Right: Oscillating
($w(a)=w_0+w_1\sin(w_2\ln(a))$), with $w_2=\frac{3}{2}\pi$.
\label{fig1}}
\end{center}
\end{figure}

\begin{figure}[htbp]
\begin{center}
\includegraphics[scale=0.9]{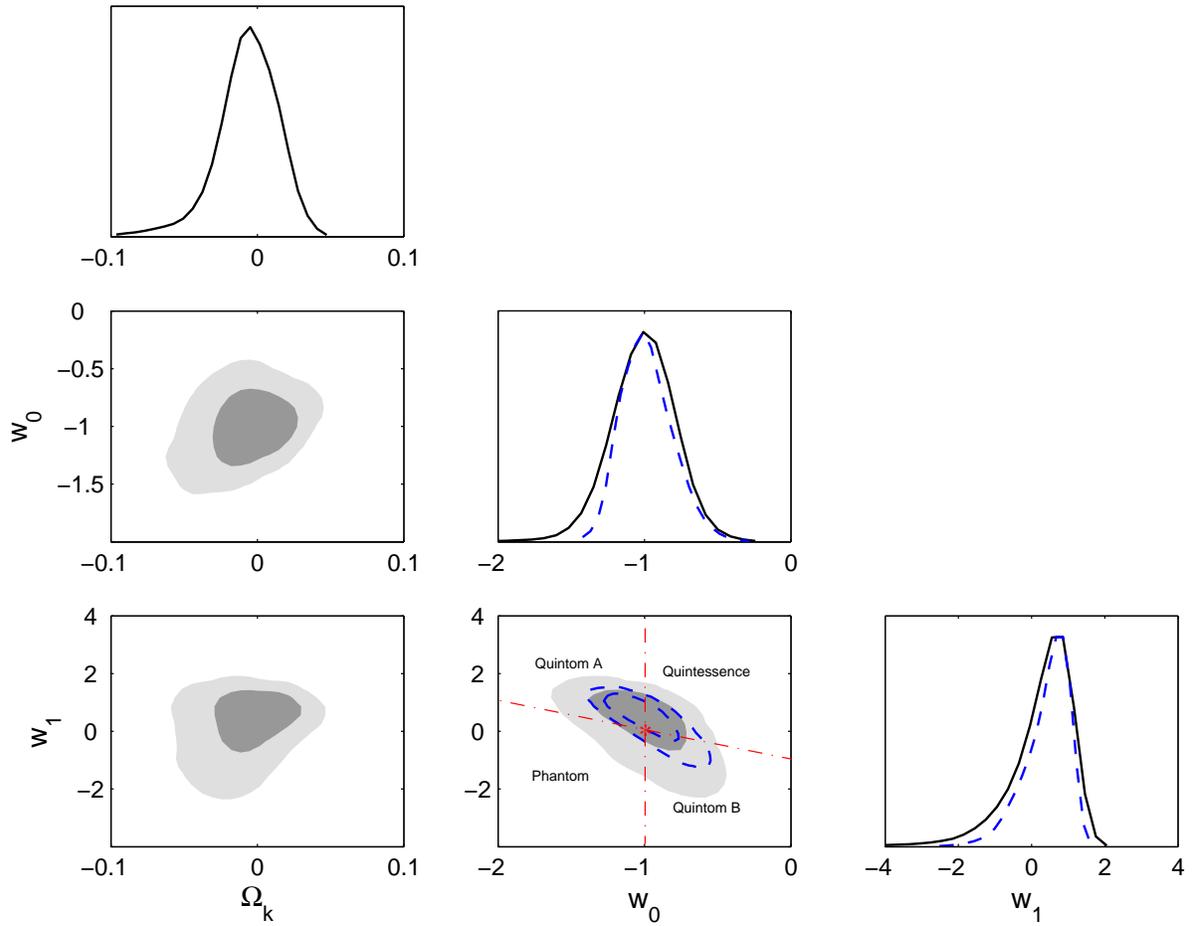}

\caption{Constraints on the curvature $\Omega_K$ and dark energy
parameters $w_0$ and $w_1$ when parameterizing DE EoS as
$w(a)=w_0+w_1(1-a)$. 1-D plots show the posterior distribution of
$\Omega_K$,$w_0$ and $w_1$ while the 2-D contour plots illustrate
their correlation. The dark and light shaded area stand for 68$\%$
and 95$\%$ C.L. respectively. Blue dashed curves denote the case
of flat universe for comparison. Different dark energy models can
be distinguished from the $w_0-w_1$ panel (See text). All these
constraints are from data of WMAP3 + SN182 + SDSS + 2dFGRS.
\label{fig2}}
\end{center}
\end{figure}

\begin{figure}[htbp]
\begin{center}
\includegraphics[scale=0.9]{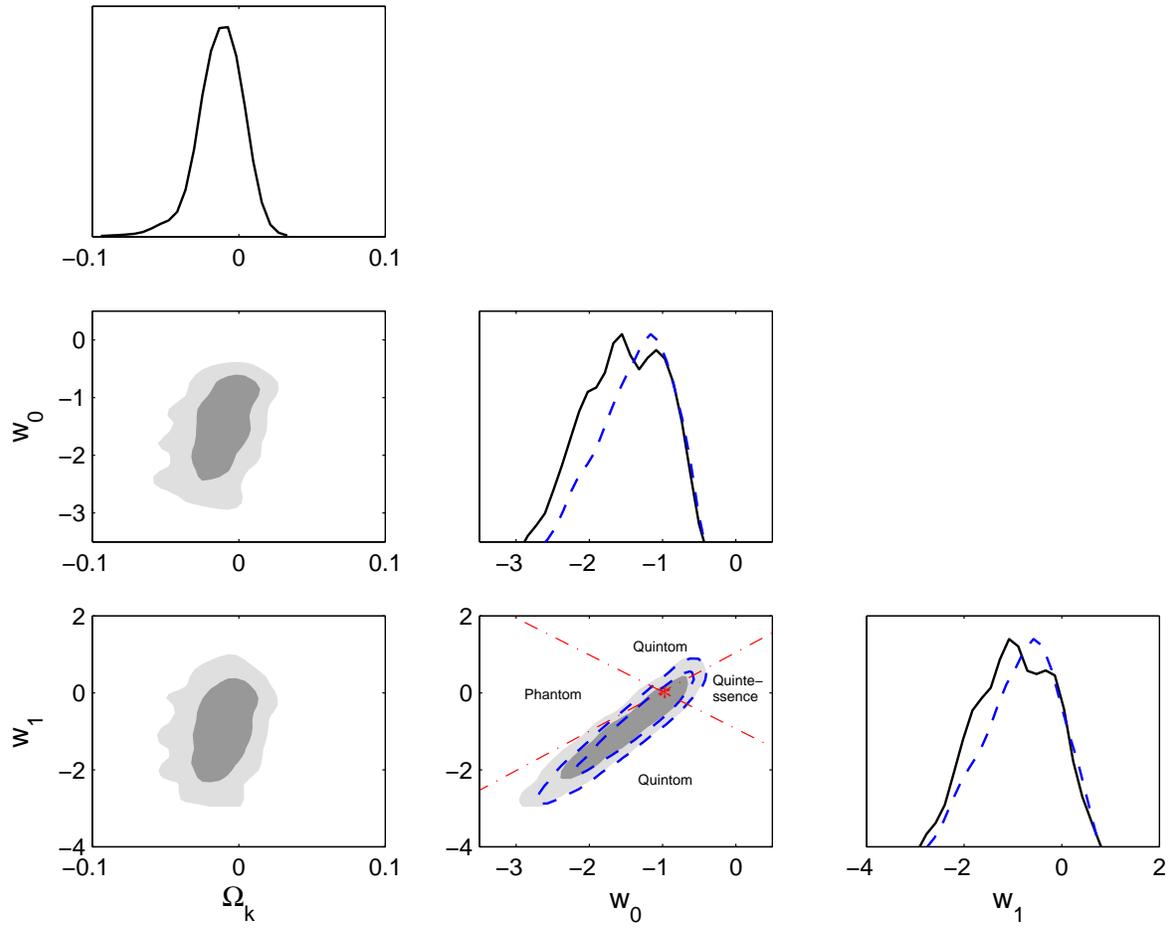}

\caption{The same graphic convention as in FIG.2 except for the
parametrization of the EoS of DE: $w(a)=w_0+w_1\sin(w_2\ln(a))$.
For simplicity, we have chosen $w_2$ to be $\frac{3}{2}\pi$. Note
the variations of the error on $\Omega_k$ for negative $\Omega_k$,
as a function of $w_1$, which reflect the oscillating feature of
this parametrization of the EoS. \label{fig3}}
\end{center}
\end{figure}

\begin{figure}[htbp]
\begin{center}
\includegraphics[scale=0.9]{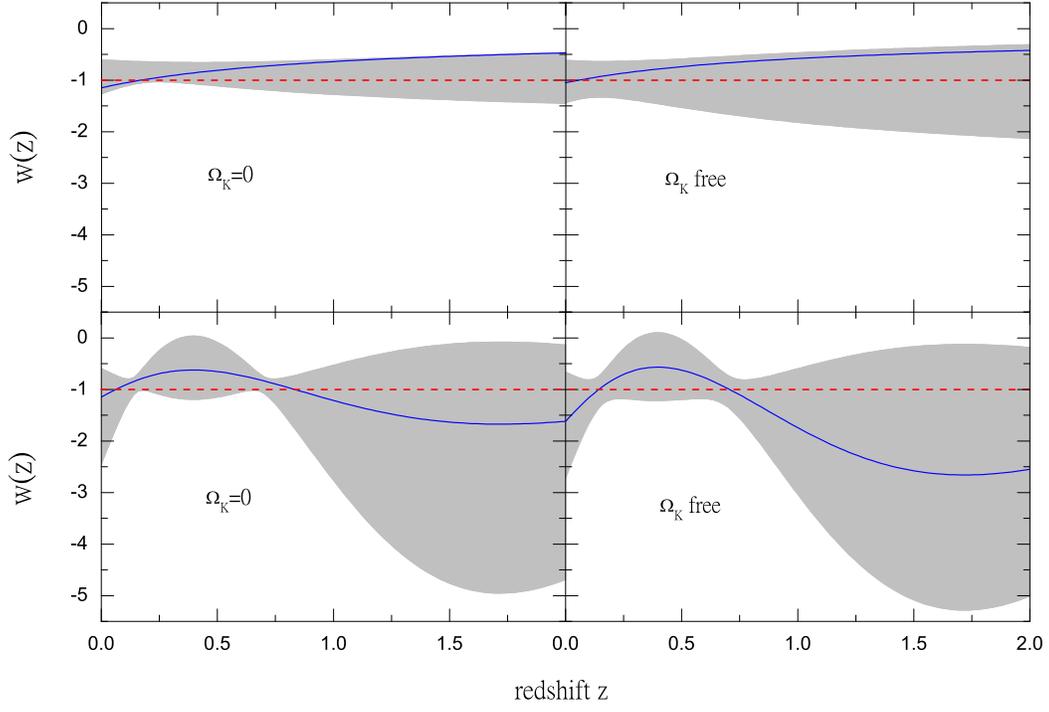}

\caption{Constraints of the time evolving of equation of state of
DE using WMAP3 + SN182 + SDSS + 2dFGRS. Upper panel:
$w(a)=w_0+w_1(1-a)$; Lower panel: $w(a)=w_0+w_1\sin(w_2\ln(a))$.
The cases of flat universe and treating $\Omega_K$ as a free
parameter are both considered as illustrated in the plot. The
solid blue lines denote the best fit models while the shaded areas
illustrate the $2\sigma$ errors. Red dashed lines show the
cosmological constant boundary. \label{fig4}}
\end{center}
\end{figure}

\end{document}